\documentstyle[12pt]{article}
\begin{document}

\baselineskip 18pt

\newcommand{\sheptitle}
{Bottom-Tau Yukawa Unification in the \\
Next-to-Minimal Supersymmetric Standard Model.}

\newcommand{\shepauthor}
{B. C. Allanach and S. F. King }

\newcommand{\shepaddress}
{Physics Department, University of Southampton\\Southampton, SO9 5NH,
U.K.}

\newcommand{\shepabstract}
{We discuss the unification of the bottom quark
and tau lepton Yukawa couplings within the framework of the
next-to-minimal supersymmetric standard model.
We compare the allowed regions of the $m_t$-$\tan \beta$ plane
to those in the minimal supersymmetric standard model,
and find that over much of the parameter space
the deviation between the predictions of two models is small,
and nearly always much less than the effect of current
theoretical and experimental uncertainties in
the bottom quark mass and the strong coupling constant.
However over some regions of parameter space top-bottom
Yukawa unification cannot be achieved.
We also discuss the scaling of the light fermion masses and
mixing angles, and show that to within current uncertainties
the results of recent texture analyses performed for the minimal
model also apply to the next-to-minimal model.}

\begin{titlepage}
\begin{flushright}
SHEP 93/94-15 \\
\end{flushright}
\vspace{.4in}
\begin{center}
{\large{\bf \sheptitle}}
\bigskip \\ \shepauthor \\ \mbox{} \\ {\it \shepaddress} \\
\vspace{.5in}
{\bf Abstract} \bigskip \end{center} \setcounter{page}{0}
\shepabstract
\end{titlepage}

It was realized some time ago that the simplest
grand unified theories (GUTs) based on SU(5)
predict the Yukawa couplings of the bottom quark
and the tau lepton to be equal at the GUT scale \cite{SU(5)},
\begin{equation}
\lambda_{b}(M_{GUT})=\lambda_{\tau}(M_{GUT})
\label{btau}
\end{equation}
where
$M_{GUT}\sim 10^{16}GeV$.
\footnote
{The relation in Eq.\ref{btau} is not exclusive to minimal $SU(5)$,
but also applies to other simple GUTs such as $SO(10)$ and $E_6$,
providing the Higgs doublets are embedded in the smallest
representations, as in the minimal $SU(5)$ model.}
Assuming the effective low energy theory below $M_{GUT}$ to be
that of the minimal supersymmetric standard model (MSSM)
the boundary condition in Eq.\ref{btau} leads to a physical
bottom to tau mass ratio $m_b/m_{\tau}$ in good agreement with
experiment
\cite{btauold}. Spurred on by recent LEP data which is
consistent with coupling constant unification,
the relation in Eq.\ref{btau} has recently been the subject of
intense scrutiny using increasingly sophisticated levels of
approximation
\cite{btaunew}, \cite{bandb}.
These analyses showed that for given
values of the strong coupling constant $\alpha_3(M_Z)$ and $m_b$,
there are only two allowed regions of $\tan \beta$
for each choice of top quark mass $m_t$: a small
$\tan \beta$ branch
and a large $\tan \beta$ branch.
\footnote{Recall that the MSSM is a two Higgs doublet model where
$H_1$ is the Higgs doublet which
gives mass to down-type quarks and charged leptons,
while $H_2$ gives mass to up-type quarks.
The superpotential contains a term $\mu H_1 H_2$,
and the ratio of the two Higgs
vacuum expectation values (VEVs) is $\tan \beta=v_2/v_1$.} However
the results are strongly
dependent on $\alpha_3(M_Z)$ and $m_b$, as well as GUT scale
threshold
effects.

The recent investigations of the relation in Eq.\ref{btau} have
focused on the MSSM.
In this paper we shall instead assume that the
effective low energy theory below the GUT scale
is the next-to-minimal
supersymmetric standard model (NMSSM) rather than the MSSM.
The NMSSM \cite{NMSSM} involves a single gauge singlet $N$
and has the $\mu$ parameter set to zero, with its effect being
replaced by terms in the superpotential like $\lambda NH_1H_2$
and $kN^3$, where $\lambda$ and $k$ are dimensionless couplings
particular to the NMSSM. The scalar component of $N$ is assumed to
develop a weak scale VEV, $x$, whose value will not directly
enter our calculations, since $N$ does not directly couple to
quarks and leptons.
The NMSSM is an alternative to the MSSM
which is equally consistent with coupling constant unification,
and is at least as well motivated as the minimal model.
Since the renormalisation group (RG) equations
of the heavy fermions involve the new couplings $\lambda$ and $k$
which may be quite large, there is no reason to expect the
allowed regions of the $m_t-\tan \beta$ plane,
consistent with Eq.\ref{btau}, to bear any resemblance
to those in the MSSM. However, as it turns out
the allowed regions in the NMSSM are quite similar to those in the
MSSM although for large values of $k (M_{SUSY})$ and $\lambda
(M_{SUSY})$ the large
$\tan \beta$ branch cannot be achieved.

Our procedure closely follows that of ref.\cite{bandb}.
We preface our discussion with a brief
summary of our approach and approximations.
For a grid of $\tan \beta$ and
$m_t$ values we determine the Yukawa couplings of the heavy fermions
at the scale $M_{SUSY}$. The three gauge couplings $g_i$ ($i=1,2,3$)
at $M_{SUSY}$ are determined by running them up from their measured
values at $M_Z$.
Having chosen values of $\lambda$ and
$k$ at $M_{SUSY}$ we run all the couplings up to
$M_{GUT}=10^{16} GeV$
using the SUSY RG equations and check if Eq.\ref{btau} is satisfied
to an accuracy of 0.1\%, for all the values of $\tan \beta$ and
$m_t$ in the grid.
Since our goal is to compare the
results of the NMSSM to those of
the MSSM at the same level of approximation,
it is sufficient to work to one loop order
in the RG equations
between $M_{SUSY}$ and $M_{GUT}$.
Similarly, we take $M_{SUSY}=m_t$ for convenience,
and ignore all low energy
threshold effects i.e. assume all SUSY partners are degenerate
with the top quark. However we shall
consider the important effects of GUT scale
threshold effects by considering
\begin{equation}
\lambda_{b}(M_{GUT})=0.9
\lambda_{\tau}(M_{GUT}) \label{btau9}
\end{equation}
rather than Eq.\ref{btau}.
In some models such as flipped SU(5) \cite{flipped}
Eq.\ref{btau9} may apply at tree-level.

The various couplings at $M_{SUSY}=m_t$ were determined
as follows. The running masses of the fermions
$m_f(\mu)$, where $f=b,\tau$ and
$\mu$ is the scale, were determined by running them up
from their mass shell values
$m_{f}(m_f)$ with effective 3~loop~QCD~$\bigotimes$~1~loop~QED
\cite{3loopqcd}, \cite{guts}.
This enables the Yukawa couplings to be determined
at $m_{t}$ by:
\begin{eqnarray}
\lambda_{t} \left( m_{t} \right) & = & \frac{\sqrt{2} m_{t}\left(
m_{t}
\right) }{v \sin\beta}\\
\lambda_{b} \left( m_{t} \right) & = & \frac{\sqrt{2} m_{b}\left(
m_{b}
\right)}{\eta_{b} v \cos \beta }\\
\lambda_{\tau} \left( m_{t} \right) & = & \frac{\sqrt{2} m_{\tau}
\left(
m_{\tau} \right)}{\eta_{\tau} v \cos\beta}. \label{Yuk}
\end{eqnarray}
 where
\begin{equation}
 \eta_{f} =\frac{m_{f} \left( m_{f} \right)}{m_{f} \left( m_{t}
\right) }
\end{equation}
and the VEV $v=\sqrt{v_1^2+v_2^2}=246$ GeV.
Clearly the Yukawa couplings at $m_t$ depend
on both $\alpha_3(M_Z)$ and $m_b(m_b)$.
In the NMSSM the additional
parameters $\lambda (m_t)$ and $k(m_t)$ are unconstrained, and may
be regarded as additional free input parameters in our analysis.
Finally the gauge couplings at $M_{SUSY}=m_t$ were determined from
some input values at $M_Z$, by using the standard model RG equations
(including 5 quark flavours and no scalar fields). The input values
were taken to be $\alpha_1(M_Z)^{-1}=58.89$,
$\alpha_2(M_Z)^{-1}=29.75$, with $\alpha_3(M_Z)=0.10-0.12$.
 Note that whereas the $m_t$ referred to here is always the
running one,
it can be related as in \cite{bandb} to the physical mass by
\begin{equation}
m_t^{phys} = m_t \left( m_t \right) \left[ 1 + \frac{4}{3 \pi}
\alpha_3 \left( m_t \right) + O \left( \alpha_3^2 \right) \right].
\end{equation}

Given the dimensionless couplings at $M_{SUSY}=m_t$,
they are then run up to $M_{GUT}=10^{16} GeV$ using the following SUSY
RG equations relevant for the NMSSM, which we obtained in a
straightforward way from ref.\cite{mandv}.
Including the full Yukawa matrices we find
\begin{eqnarray}
\frac{\partial U}{\partial t} &=& \frac{U}{16 \pi^{2}} \left[ 3
\mbox{Tr}
\left( U U^{\dagger} \right) + 3 U^{\dagger} U +
D^{\dagger} D + \lambda^{2}
- \left( \frac{13}{15} g_{1}^{2} + 3g_{2}^{2}
+ \frac{16}{3} g_{3}^{2} \right )\right] \nonumber \\
\frac{\partial D}{\partial t} &=& \frac{D}{16 \pi^{2}} \left[
\mbox{Tr}
\left( 3D D^{\dagger} + E E^{\dagger} \right) + U^{\dagger} U + 3
D^{\dagger} D + \lambda^{2} - \left( \frac{7}{15} g_{1}^{2} + 3
g_{2}^{2} + \frac{16}{3} g_{3}^{2} \right) \right] \nonumber \\
\frac{\partial E}{\partial t} &=& \frac{E}{16 \pi^{2}} \left[
\mbox{Tr}
\left( 3 D D^{\dagger} + E E^{\dagger} \right) + 3
E^{\dagger} E + \lambda^{2}
- \left( \frac{9}{5} g_{1}^{2} + 3 g_{2}^{2} \right)
\right] \nonumber \\
\frac{\partial \lambda}{\partial t} &=& \frac{\lambda}{16 \pi^{2}}
\left[ \mbox{Tr} \left( 3 U U^{\dagger} + 3 D D^{\dagger} + E
E^{\dagger}
\right) + 4 \lambda^{2} + 2 k^{2} - \left( \frac{3}{5} g_{1}^{2} + 3
g_{2}^{2} \right) \right] \nonumber \\
\frac{\partial k}{\partial t} &=& \frac{6 k}{16 \pi^{2}} \left[ k^{2} +
\lambda^{2} \right], \label{RG1}
\end{eqnarray}
where $U$, $D$ and $E$ are the up, down and charged lepton Yukawa
matrices
respectively and
the GUT normalisation convention of $g_{1}$ has been used.
Dropping small Yukawa couplings Eq.\ref{RG1} reduces to
\begin{eqnarray}
16 \pi^{2} \frac{\partial \lambda_{t}}{\partial t} &=& \lambda_{t}
\left[ 6\lambda_{t}^{2} +  \lambda_{b}^{2} +  \lambda^{2} -
\left( \frac{13}{15}
g_{1}^{2} + 3g_{2}^{2} + \frac{16}{3}g_{3}^{2} \right)
\right] \nonumber \\
16 \pi^{2} \frac{\partial \lambda_{b}}{\partial t} &=&
\lambda_{b} \left[ 6
\lambda_{b}^{2} + \lambda_{\tau}^{2} + \lambda_{t}^{2} + \lambda^{2} -
\left(
\frac{7}{15} g_{1}^{2} + 3g_{2}^{2} +
\frac{16}{3} g_{3}^{2} \right) \right] \nonumber \\
16 \pi^{2} \frac{\partial \lambda_{\tau}}{\partial t} &=&
\lambda_{\tau}
\left[ \lambda_{\tau}^{2} + 3 \lambda_{b}^{2}  + \lambda^{2}
- \left( \frac{9}{5} g_{1}^{2}
+ 3g_{2}^{2} \right) \right] \nonumber \\
16 \pi^{2} \frac{\partial \lambda}{\partial t} &=& \lambda
\left[ 4 \lambda^{2} + 2 k^{2} + 3\lambda_{\tau}^{2} + 3 \lambda_{b}^{2}
+ 3 \lambda_{t}^{2}
- \left(
\frac{3}{5} g_{1}^{2} + 3g_{2}^{2} \right) \right] \nonumber \\
16 \pi^{2} \frac{\partial k}{\partial t} &=& 6 k \left[ \lambda^{2}+
k^{2}
\right]. \label{krg}
\end{eqnarray}

Our results are displayed in Fig.1 as
contours in the $\tan \beta - m_t$ plane consistent
with Eq.\ref{btau}. We take
$\alpha_3(M_Z)=0.11$, $m_b=4.25 GeV$ and the NMSSM parameters
$\lambda (m_t)$ and $k(m_t)$ as indicated.
The MSSM contour is shown for comparison and is
indistinguishable from the NMSSM contour with
$\lambda \left( m_{t} \right) =0.1$ and $k \left( m_{t} \right) =0.5$.
In fact our plot for the MSSM based on 1-loop RG equations
is very similar to the 2-loop result in ref.\cite{bandb}.
The deviation of the NMSSM contours from the MSSM contour
depends most sensitively on $\lambda(m_t)$ rather than $k(m_t)$.
Two of the contours are shortened due to either $\lambda$ or $k$
blowing up at the GUT scale.
For $\lambda \left(
m_t \right) =0.5$, $k \left( m_t \right) =0.5$, no points in
the $m_{t} - \tan \beta $ plane are consistent with Eq.\ref{btau}
Yukawa unification, while for $\lambda \left(
m_t \right) =0.1$, $k \left( m_t \right) =0.1-0.5$
the contours are virtually indistinguishable from the MSSM contour.
In general we find that for
any of the current experimental limits on $\alpha_3$ and $m_b$, the
maximum value of $\lambda (m_t)$ or $k (m_t)$ is $\sim 0.7$ for a
perturbative solution to Eq.\ref{btau}.

In Figs.2 and 3 we examine the effect of the experimental
uncertainties in $m_b$ and $\alpha_3(M_Z)$
as well as the theoretical uncertainties
parameterised by Eq.\ref{btau9}.
In Fig.2, for a rather low value of $\alpha_3(M_Z)=0.10$,
the region between the two
solid lines respects Eq.\ref{btau}
in the NMSSM with $\lambda (m_t)=0.5$, $k=0.1$,
and $m_b=4.1-4.4GeV$, while the dashed
line satisfies Eq.\ref{btau9} for $m_b=4.4GeV$.
The corresponding dashed
line which satisfies Eq.\ref{btau9} for $m_b=4.1GeV$
occurs for $m_t<100GeV$ and so is to the left of the figure.
Clearly for this value of strong coupling, the error in the bottom
quark mass turns the sharp contours in Fig.1 into thick bands,
leading to no prediction of $\tan \beta$ for $m_t>150GeV$.
In Fig.3, for a higher value of
$\alpha_3(M_Z)=0.12$, the same uncertainties have a much
less severe effect than in Fig.2, and larger values of the top
quark mass are permitted.
However the moral of Figs.2 and 3 is
clear: the sharp contours of Fig.1 must not be taken too seriously
given the present experimental and theoretical uncertainties.

So far we have discussed the heavy third family fermion masses only.
Let us now extend our discussion of the NMSSM to include the light
fermion masses and mixing angles. It is well known in the MSSM that
to one loop order in the RG equations,
the running of the physically
relevant Yukawa eigenvalues and mixing angles can be expressed in
simple terms as shown below,
\begin{eqnarray}
\left( \frac{\lambda_{u,c}}{\lambda_{t}} \right)_{M_{SUSY}} & = &
\left( \frac{\lambda_{u,c}}
{\lambda_{t}} \right)_{M_{GUT}} e^{3I_{t} + I_{b}} \label{start}
\nonumber \\
\left( \frac{\lambda_{d,s}}{\lambda_{b}} \right)_{M_{SUSY}} & = &
\left( \frac{\lambda_{d,s}}
{\lambda_{b}} \right)_{M_{GUT}} e^{3I_{b} + I_{t}} \nonumber \\
\left( \frac{\lambda_{e, \mu}}{\lambda_{\tau}} \right)_{M_{SUSY}} & = & \left(
\frac{\lambda_{e, \mu}}
{\lambda_{\tau}} \right)_{M_{GUT}} e^{3I_{\tau}} \nonumber \\
\frac{ \mid V_{cb} \mid _{M_{GUT}}}{\mid V_{cb} \mid
_{M_{SUSY}} }
& = & e^{I_{b}+I_{t}} , \label{gutsusy}
\end{eqnarray}
with identical scaling behaviour to $V_{cb}$ of $V_{ub}$,
$V_{ts}$, $V_{td}$, where
\begin{equation}
I_{i}=\int _{\ln M_{SUSY}}^{\ln {M_{GUT}}} \left( \frac{\lambda_{i} \left( t
\right) }{4 \pi} \right)^2 dt. \label{iint}
\end{equation}
To a consistent level of approximation $V_{us}$, $V_{ud}$, $V_{cs}$,
$V_{cd}$, $V_{tb}$, $\lambda_{u}$/$\lambda_{c}$,
$\lambda_{d}$/$\lambda_{s}$ and $\lambda_{e}$/$\lambda_{\mu}$ are RG
invariant.
The CP violating quantity J scales as $V_{cb}^{2}$.
The Eqs. \ref{gutsusy}, \ref{iint} also apply to the NMSSM since
the extra $\lambda$ and $k$ paramters
cancel out of the RG equations in a similar way to the gauge contributions
as can easily be seen from Eq.\ref{RG1}.
The only difference to these physically relevant
quantities is therefore contained in $I_{\tau}$, $I_{b}$ and $I_{t}$.

In Figs.4-6 we plot the values of $I_{\tau}$, $I_{b}$ and $I_{t}$
as evaluated along the contours in Figs.1-3 which satisfy Eq.\ref{btau}.
Fig.4 illustrates the difference in the $I_{i}$ quantities
between the MSSM and the NMSSM for the contour in Fig.1 corresponding
to $\lambda (m_t)=0.5$, $k(m_t)=0.1$. The NMSSM results are the upper
lines of each pair, and it is clear that the deviation between the two
models is small. In Fig.5 we plot the three $I_i$ integrals
for the NMSSM along the two solid contours of Fig.2,
corresponding to a low value of $\alpha_3(M_Z)=0.10$
and a range of $m_b=4.1-4.4GeV$. The experimental uncertainty
in $m_b$ shown in Fig.5 clearly swamps the theoretical difference
between the values of $I_{b}$ and $I_{t}$ for the
NMSSM and the MSSM shown in Fig.4, for this choice of
$\alpha_3(M_Z)$. In Fig.6 we plot the three $I_i$ integrals
along the two solid contours of Fig.3,
corresponding to a higher value of $\alpha_3(M_Z)=0.12$
and the same range of $m_b$. Again the large deviations
in $I_{b}$ and $I_{t}$ due to the error in $m_b$ swamp the theoretical
differences between the two models. However, the values of $I_{\tau}$
in Figs.5,6 are quite robust, and the theoretical deviation
in $I_{\tau}$ shown in Fig.4 is significant, and may play
an important role in distinguishing between the two models.

We emphasise that the results of the $I_i$ integrals
shown in Figs.4-6 play a key role in determining
the entire fermion mass spectrum via the scaling relations
shown in Eq.\ref{gutsusy}. The small deviation between the NMSSM and the MSSM
results compared to the experimental uncertainties,
means that the recent GUT scale
texture analyses of the quark mass matrices which were performed
for the MSSM are equally applicable to the NMSSM.
For example, the recent Ramond, Roberts and Ross (RRR) \cite{RRR}
texture analysis is also based upon Eq.\ref{btau} and
assumes a Georgi-Jarlskog
(GJ) \cite{GJ},
\cite{DHR} ansatz
for the charged lepton Yukawa matrices, although their results in the
quark sector
are insensitive to the lepton sector. It is clear that all the RRR
results are immediately applicable to the NMSSM without further
ado since the only difference between the two models
enters through the scaling integrals $I_i$ whose deviation
we have shown to be
negligible compared to the experimental errors.

Finally in Fig.7 we examine the question of full top-bottom-tau
Yukawa unification in the NMSSM. Third family Yukawa unification
has recently been studied in some detail in the MSSM \cite{yuk}
and occurs theoretically in minimal $SO(10)$ \cite{so10}
and $SU(4)~\otimes~SU(2)^2$ \cite{422} models.
Fig.7 shows the values of the Yukawa couplings evaluated at the GUT scale
for values of $\tan \beta$ along two of the contours in Fig.1
corresponding to the NMSSM with $\lambda (m_t)=0.5$ and $k(m_t)=0.1,0.4$.
The tau Yukawa coupling is of course equal to the bottom Yukawa coupling
at the GUT scale and so is not labelled explicitly in this figure.
In Fig.1 it was observed that
the high $k$ lines are shorter
than the others and consequently the large $\tan \beta$ regions cannot be
achieved. The reason is clear from Fig.7, which shows that $k(M_{GUT})$
blows up for large $\tan \beta$. This is
because, as can be seen from Eq.\ref{krg}, $k$ scales
quickly and readily becomes nonperturbative for $k(m_t)>0.5$.
The stunted lines in Fig.7 mean that, for
$k(m_t)=0.4$ and the other parameters as assumed in the figure,
top-bottom Yukawa unification cannot be achieved.
The longer lines in Fig.7 show that, for $k(m_t)=0.1$
and all the other parameters unchanged,
larger values of $\tan \beta$ may be achieved
and top-bottom unification, which occurs for $\tan \beta \approx 50$
for this set of parameters, is once again possible.

In conclusion,
we have discussed the unification of the bottom quark
and tau lepton Yukawa couplings within the framework of the
NMSSM. By comparing the allowed regions of the $m_t$-$\tan \beta$ plane
to those in the MSSM
we find that over much of the parameter space
the deviation between the predictions of two models
which is controlled by the parameter $\lambda$ is small,
and always much less than the effect of current
theoretical and experimental uncertainties in
the bottom quark mass and the strong coupling constant.
We have also discussed the scaling of the light fermion masses and
mixing angles, and shown that to within current uncertainties,
the results of recent quark texture analyses \cite{RRR}
performed for the minimal
model also apply to the next-to-minimal model. There are however
two distinguishing features of the NMSSM. Firstly, the scaling
of the charged lepton masses will be somewhat different,
depending on $\lambda$ and $k$. Although this will not affect
the quark texture analysis of RRR, it may affect the success of the
GJ ansatz \cite{GJ}, \cite{DHR} for example. Secondly, the larger
$\tan \beta$ regions may not be accessible in the NMSSM
for large values of $\lambda$ and $k$, so that full Yukawa unification
may not be possible in this case.

\newpage

{\bf \Large Figure Captions}
\vspace{0.25in}

{\bf Figure 1.}
Contours in the $m_t-\tan \beta$ plane
over which the bottom-tau unification condition
Eq.\ref{btau} is satisfied in
the MSSM and the NMSSM.
Central values of $\alpha_{3}(M_Z) = 0.11$
and $m_{b}= 4.25$ GeV are assumed, and NMSSM contours for various
$\lambda (m_t)$ and $k(m_t)$ values are shown.

\vspace{0.25in}

{\bf Figure 2.}
Contours in the $m_t-\tan \beta$ plane
over which Eq.\ref{btau} is satisfied in
the NMSSM with $\lambda (m_t)=0.5$ and $k(m_t)=0.1$, for
a low value of $\alpha_{3} \left( M_{Z} \right) = 0.10$.
The region between the two
solid lines is for exact bottom-tau unification,
with $m_b=4.1-4.4GeV$. The dashed line
satisfies  $\lambda_{b}=0.9 \lambda_{\tau}$ as in Eq.\ref{btau9}
for $m_b=4.4GeV$.

\vspace{0.25in}

{\bf Figure 3.}
Contours in the $m_t-\tan \beta$ plane
over which Eq.\ref{btau} is satisfied in
the NMSSM with $\lambda (m_t)=0.5$ and $k(m_t)=0.1$, for
a high value of $\alpha_{3} \left( M_{Z} \right) = 0.12$.
The region between the two
solid lines is for exact bottom-tau unification,
with $m_b=4.1-4.4GeV$. The region between the two dashed lines
satisfies $\lambda_{b}=0.9 \lambda_{\tau}$ as in Eq.\ref{btau9}
for $m_b=4.1-4.4GeV$.

\vspace{0.25in}

{\bf Figure 4. }
The $I_f$ integrals defined in the text as evaluated along
two of the bottom-tau unification
contours in Fig.1. As in Fig.1, $\alpha_{3}(M_Z) = 0.11$
and $m_{b}= 4.25$ GeV.
The pairs of lines shown in this figure
correspond to the
MSSM (lower lines) and the NMSSM with
$ \lambda (m_t)=0.5$, $k(m_t)=0.1$ (upper lines).

\vspace{0.25in}

{\bf Figure 5. }
The $I_f$ integrals defined in the text as evaluated along
the two exact bottom-tau unification
contours of Fig.2. As in Fig.2,
$\alpha_{3}(M_{Z}) = 0.10$,
$\lambda_{b}(m_t)=0.5$, $k(m_t)=0.1$.
The shorter lines in this figure correspond
to $m_b=4.1GeV$, the longer lines to $m_b=4.4GeV$.

\vspace{0.25in}

{\bf Figure 6.}
The $I_f$ integrals defined in the text as evaluated along
the two exact bottom-tau unification
contours of Fig.3. As in Fig.3,
$\alpha_{3}(M_{Z}) = 0.12$,
$\lambda_{b}(m_t)=0.5$, $k(m_t)=0.1$.
The shorter lines in this figure correspond
to $m_b=4.4GeV$, the longer lines to $m_b=4.1GeV$.

\vspace{0.25in}

{\bf Figure 7.}
The various NMSSM couplings evaluated
at the GUT scale, corresponding to the two
bottom-tau unification contours in Fig.1
with $\lambda (m_t) =0.5$.
The longer lines in this figure
are from the $ k(m_t) =0.1$ line, and exhibit
top-bottom unification for $\tan \beta \approx 50$.
The shorter ones are from the $ k(m_t) = 0.4$ curve,
and show that top-bottom unification cannot be achieved since
$k$ blows up before a sufficiently large $\tan \beta$ can
be achieved.

\end{document}